\documentclass[preprint2]{aastex}
\def\f{\frac}
\def\p{\partial}
\def\al{&\!\!\!\!}

\slugcomment{Not to appear in Nonlearned J., 45.}

\shorttitle{} \shortauthors{}

\begin{document}


\title{Transverse Oscillations of Longitudinally Stratified Coronal Loops System}

\author{N. Fathalian\altaffilmark{1} and H. Safari\altaffilmark{2}}
\affil{$^1$Institute for Advanced Studies in Basic Sciences, P. O.
Box 45195-1159, Zanjan, Iran\\$^2$Department of Physics, Zanjan
University, P. O. Box 45195-313, Zanjan, Iran}
\email{fathalian@iasbs.ac.ir, safari@znu.ac.ir}
%
%

\begin{abstract}
The collective transverse coronal loop oscillations seem to be
detected in the observational studies. In this regard, Luna et al.
(2009, ApJ, 692, 1582) modeled the collective kinklike normal
modes of several cylindrical loops system using the T-matrix
theory.

This paper investigates the effects of longitudinal density
stratification along the loop axis,
 on the collective kinklike modes of system of
 coronal loops. The coronal loops system
 is modeled as cylinders of parallel flux tubes, with two ends of each loop
at the dense photosphere. The flux tubes are considered as uniform
magnetic fields, with
 stratified density along the loop axis
which changes discontinuously at the lateral surface of each
cylinder. The MHD equations are reduced to solve a set of two
coupled dispersion relations for frequencies and wave numbers, in
the presence of stratification parameter. The fundamental and
first overtone frequencies and longitudinal wave numbers are
computed. The previous results are verified for unstratified
coronal loops system.

Finally we would conclude that increased longitudinal density
stratification parameter will result in increase of the
frequencies. The frequencies ratios, first overtones to
fundamentals, are very sensitive functions of density scale
height parameter. Therefore, stratification should be included in
dynamics of coronal loops systems. For the unstratified
 coronal loops system, these ratios are the same as monoloop
ones.
\end{abstract}
\section{Introduction}
The high-resolution observations of TRACE, SoHO, Yohkoh, etc
provided us with detection of coronal waves (e.g., Aschwanden et
al. 1999a, b, 2002; Nakariakov et al. 1999; Schrijver \& Brown
2000 and Verwichte et al. 2004). Throughout the development of the
observations, transverse and longitudinal oscillations have been
studied.
 The coronal seismology techniques allow the information to be extracted from
observations of oscillatory phenomena and the results to be
interpreted, using theoretical models (e.g., Edwin \& Roberts
1983; Roberts et al. 1984; Goossens et al. 1992).

In the recent observational data, periods, phases, damping times,
and mode profiles for coronal loops are reported by Verwichte et
al. (2004) and De Moortel \& Brady (2007). Expectedly, the results
differ from those based on simplified theoretical models. To be
more realistic, several features may be added to this simple
model, such as the presence of magnetic twist and shells,
field-aligned flows, the role of line-tying effects, loop
curvature, coronal leakage, etc.

Andries et al. (2005a,b) pointed out the effect of longitudinally
density stratification as an important feature on coronal loop
model and oscillations.  Andries et al. (2005a, b),
 Donnelly et al. (2006), Dymova \& Ruderman (2006), McEwan et al. (2006),
 Erd\'{e}lyi, \& Verth (2007), Safari et al. (2007), and Ruderman et al.
 (2008), used the frequencies ratio, $P_1/P_2$, as a siesmological tool
for estimating the solar atmosphere density scale height.
Recently, Andries et al. (2009) and Aschwanden (2009), reviewed
the details of this topic.

Another added feature is the collective nature of the
oscillations. This idea comes from the bundles or arcades of loops
observed in active regions. As observational instances, Schrijver
and Brown (2000) observed antiphase transverse oscillations of
adjacent loops and Verwichte et al. (2004) reported phase and
antiphase motions, in a post-flare arcade. The exact structures of
the active regions coronal loops have not been determined yet.
Actually, we don't know exactly, weather those structures are
monolithic or multistranded. Multistranded model assumes that
each loop is composed of miniloops -several tens or hundreds of
strands (Klimchuk 2006). In the context of non individual flux
tube oscillations, the propagation of fast waves in two slabs
(Murawski 1993; Murawski \& Roberts 1994), and the oscillations
of the prominence threaded structure (D\'{i}az et al. 2005) have
been studied.

Gruszecki et al. (2006), considered impulsively generated
oscillations in a 2D model of a curved solar coronal arcade loop
that consists of up to 5 strands of dense plasma. Pascoe et al.
(2007), studied the effect of fine multishell structuring on the
resonant periods of global sausage fast MHD oscillations of
straight magnetic slab model of coronal loops. They indicated
that the resonant properties of long-wavelength sausage standing
modes are not sensitive to fine structuring.

Luna et al. (2009), studied the collective kinklike normal modes
of loops, which are set with different physical and geometrical
properties. They used the scattering theory, T-matrix, as
extended by Waterman \& Truell (1961) and Ramm (1986). Luna et
al. (2009), concluded that loops with similar kink frequencies,
oscillate collectively with a frequency slightly different from
that of the individual kink mode. Otherwise, a loop with different
 kink frequency oscillates individually with its own frequency.
The kink frequencies of neighboring loops with similar densities
are coupled.  Luna et al. (2010), investigated the transverse
oscillations of a multi-stranded coronal loop, composed of several
parallel cylindrical strands. They concluded that, the presumed
internal fine structure of a loop influences its transverse
oscillations and so presumable multi-stranded coronal loop
transverse dynamics cannot be properly described by those of an
equivalent monolithic loop.

Here, we generalize  Luna et al. (2009) method to include
 longitudinal density stratification in collective oscillations of a coronal loops system.
 We suppose parallel cylindrical flux tubes, with their ends at
the photosphere and with a relatively small curvatures. The
cylinders are assumed to have no initial material flow, to be
pervaded by uniform magnetic fields along their axis,
 and to have negligible gas pressure (zero-$\beta$ approximation).
We use a single PDE equation for $z-$component of perturbed
magnetic filed as derived by Safari et al. (2007), for studying
the oscillations of a single isolated longitudinally stratified
 thin coronal loop. Assuming stratified flux tubes system, the
radial and longitudinal parts of this equation are separated. The
radial part is solved based on T-matrix theory extended by Luna et
al. (2009) and the longitudinal part is investigated similar as
Safari et al. (2007).  We derive a set of two coupled dispersion
relations for oscillations frequencies and longitudinal wave
numbers.  In the case of unstratified flux tubes system, our
approach is similar to that of Luna et al. (2009), which have
reduced the MHD equations from the beginning to accommodate
system of uniform flux tubes. We calculate the eigenvalues and
eigenfunctions of the normal modes of the stratified model.

This paper's layout is as follows: The physical model and the set
of two dispersion relations are treated in Secs \ref{physmod} and
\ref{sec3}, respectively. The numerical results and conclusions
are presented in Sec. \ref{numeric}.
\section{Equilibrium model and equations of motions}\label{physmod}
\begin{figure}
 \includegraphics{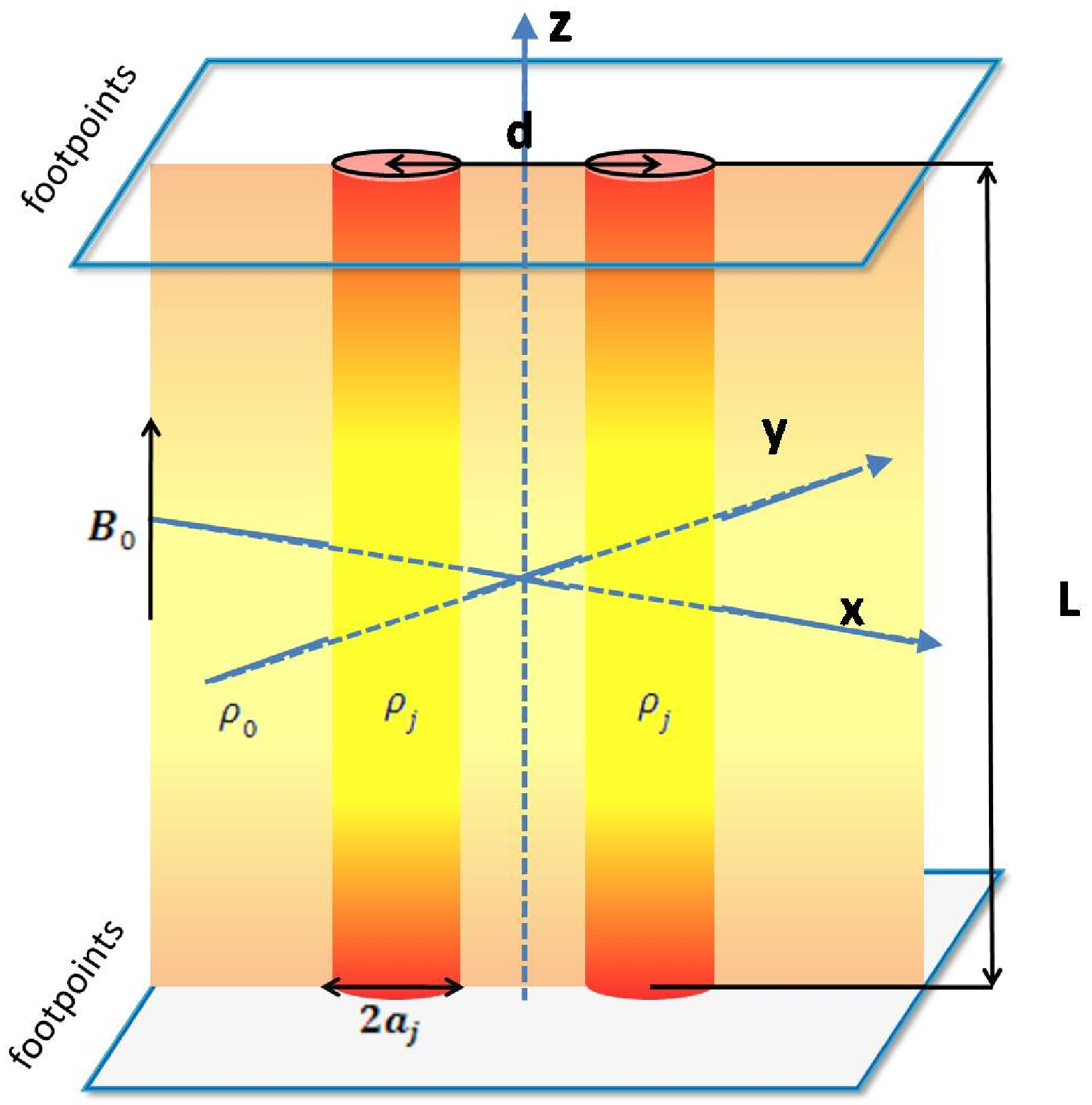}
         \vspace{6cm}
       \caption{A sketch of a system of two straight flux tubes which are
       presented in the coronal medium. Interior  and exterior densities vary along the
cylinders axis and are symmetric about the midpoints. The magnetic
field is uniform along the $z-$axis.
            }
            \label{fig0}
\end{figure}

We use a system of cylindrical coordinates, $(r,\phi,z)$. The
equilibrium configuration of coronal loops systems is modeled as a
set of cylinders with their axis along the $z$-coordinate. Each
loop characterized as, $j$, has the length of $L$, radius $a_j$,
internal density, $\rho_{j}(r,\phi,z,\epsilon)$, and centered at
$\mathbf{r}_j=x_j\mathbf{e_x}+y_j\mathbf{e_y}$, in $xy$-plane,
Fig \ref{fig0}.
The equilibrium magnetic field is uniform,
 $\mathbf{B}=B_0\mathbf{e}_z$.
The stratified density of each loop, $j$, is assumed to be
\begin{eqnarray}\label{density}
\rho_j(r,\phi,z,\epsilon)\al=\al
\rho_{j}(\epsilon) f(\epsilon,z),
~|\mathbf{r}-\mathbf{r}_j|\leq a_j,\nonumber\\
\al=\al\rho_{0}(\epsilon)f(\epsilon,z),
~\rm{~for~external~of~loops},\nonumber\\
\tilde{\rho}=\f{\rho_{j}(\epsilon)}{\rho_j(0)}\al=\al\f{\rho_{0}(\epsilon)}{\rho_0(0)}=\frac{1}{\int^{L}_0{f(\epsilon,z)dz}},
~\epsilon=\f{L}{H}
\end{eqnarray}
where, $\epsilon$ is the stratification parameter and $H$ is the
density scale height. The internal and external footpoint
densities are $\rho_{j}(\epsilon)$ and $\rho_{0}(\epsilon)$ for
stratified loops and $\rho_{j}(0)$ and $\rho_{0}(0)$ for
unstratified loops. See Safari et al. (2007) and Fathalian et al.
(2010).

The set of linearized MHD equations are reduced as a single PDE
for $z-$component of perturbed magnetic field, $b_z$, as
\begin{equation}\label{nablaz}
\nabla^2 b_z(r,\varphi,z)-\frac{\omega^2}{v_A^2(r,\varphi,z,\epsilon)}b_z(r,\varphi,z)=0,
\end{equation}
where, $v_A=B_0/\sqrt{4\pi\rho(r,\varphi,z,\epsilon)}$, the
Alfv\'{e}n velocity, is the step function of $r$ and $\varphi$.
Note that, we used Fourier transform for all perturbed
quantities,  $\exp{(i\omega t)}$. See Safari et al. (2007) for
derivation details.

Following as Safari et al. (2007),  using Eq. (\ref{nablaz}) and
$b_z(r,\varphi,z)=\psi(r,\varphi)Z(z)$ we find
\begin{eqnarray}\label{in1}
\nabla^2_\bot\psi^{j}+k^2_{j}\psi^{j}=0
,~j=1,2,...,N,\\
(\f{d^2}{dz^2}-k^2_{j})Z^{j}+
\f{\omega^2}{v_{A_{j}}^2(z,\epsilon)}Z^{j}=0,\label{in2}
\end{eqnarray}
for internal of the loops,
in which
\begin{eqnarray}\label{kj}
\al\al k^2_{j}=\f{\omega^2}{v_{A_{j}}^2(z,\epsilon=0)}-k_z^2,\\
\al\al \nabla^2_\bot=\frac{1}{r}\f{\p}{\p r}(r \f{\p}{\p
r})+\f{1}{r^2}\f{\p ^2}{\p \varphi^2},\nonumber
\end{eqnarray}
where, $N$ and $k_z^2$ are number of loops and a constant,
respectively. Similar relations can be obtained for external part
of the loops by replacing "$j$" with "$0$".

 The changes in total
pressure should be continuous at the tube lateral surface. On
account of the zero-$\beta$ approximation and constancy of
equilibrium magnetic field, $B$, this reduces to the requirement
of the continuity of perturbed magnetic filed $b_z$. Thus,
\begin{eqnarray}\label{b}
\al\al\psi^j(r,\varphi)=\psi^0(r,\varphi)|_{\mathbf{r}=\mathbf{r}_j+\mathbf{a}_j},~j=1,2,...,N,
\nonumber\\
\al\al
Z^{j}(z)=Z^{0}(z)|_{\mathbf{r}=\mathbf{r}_j+\mathbf{a}_j}=Z(z).
\end{eqnarray}

Using the boundary conditions, Eq. (\ref{b}), and Eq. (\ref{in1})
and (\ref{in2}), containing the equations for external part we
get to
\begin{eqnarray}\label{d2z}
\f{d^2 Z}{dz^2}+A\omega^2(\tilde{\rho} f(\epsilon,z)-1)Z+k_z^2 Z=0,
\end{eqnarray}
where the constant $A$ is defined by,
\begin{eqnarray}\label{A}
(1+N)A=\sum_{j=1}^N
\f{1}{v_{A_{j}}^2(0,z)}+\f{1}{v_{A_{0}}^2(0,z)}.\nonumber
\end{eqnarray}

Hereafter, we restrict $z$ to the interval $[0,L/2]$, because of
symmetry of each loop about its midpoint, $z=L/2$. For
exponentially stratified plasma, $f(\epsilon,z)=\exp({-\epsilon
z/L})$. So, the solution of Eq. (\ref{d2z}) is:
\begin{eqnarray}\label{Z}
\al\al \hspace{-0.5cm}Z(z)= C_1 J_{\nu}
(\f{2\tilde{\omega}e^{-\f{1}{2}\epsilon z}}{\epsilon})+ C_2
Y_{\nu} (\f{2\tilde{\omega}e^{-\f{1}{2}\epsilon
z}}{\epsilon}),~~~~
\end{eqnarray}
in which
$\nu=-\f{2\sqrt{\tilde{\omega}^2/\tilde{\rho}-k_z^2}}{\epsilon}$
and $\tilde{\omega}^2=A\tilde{\rho}\omega^2$.
\section{Dispersion relations }\label{sec3}
\subsection{Dispersion relations of radial part, Eq. (\ref{in1})}\label{disp-r}
Following Luna et al. (2009), for a system of $N$ coronal loops,
Eq. (\ref{in1}) could be solved by applying T-matrix theory. The
T-matrix method states that the net external field is composed of
addition of outgoing scattered waves (Bogdan and Cattaneo 1989)
and the wave scattered by the $j-$th loop is the outcome of a
response of the external field minus the contribution of the
mentioned loop. The following linear algebraic system of
equations for the complex coefficients $\alpha_m^j$ can be
obtained
\begin{eqnarray}\label{alpha}
\al\al\hspace{-.5cm} \sum_{i\neq j}^{N}\sum_{n=-\infty}^{n=\infty}
\alpha_n^i T_{nn}^i  H^{(1)}_{n-m}(k_0|\mathbf{r}_j-\mathbf{r}_i|)e^{i(n-m)\varphi_{ji}}\nonumber\\
\al\al\hspace{0cm}+\alpha_m^j=0,~j=1,\cdots,N,~m=1,\cdots,m_{t},
\end{eqnarray}
in which $\alpha_m^j$ are the expansion coefficient of order $m$,
 $k_0$ is the wave number in the external medium, $|\mathbf{r}_j-\mathbf{r}_i|$ and
$\varphi_{ji}$ are the distance and the angle formed by the center
of the $i-$th loop with respect to the center of the $j-$th loop.
$H^{(1)}_m$ are the Hankel functions of the first kind. The matrix
diagonal elements, $T^j_{mm}$, of the operator $T^j$ are
\begin{eqnarray}\label{T}
\al\al T^j_{mm} =\\
\al\al\f{k_j^2 k_0J_m(k_ja_j)J'_m(k_0a_j)-k_0^2 k_{j}J'_m(k_ja_j)J_m(k_0a_j)}
{k_0^2 k_{j}H^{(1)}_m(k_0a_j)J'_m(k_ja_j)-k_j^2 k_0H'^{(1)}_m(k_0a_j)J_m(k_ja_j)}.\nonumber
\end{eqnarray}
The sign "$'$" denotes the derivative of function in respect of
argument. For more details see Luna et al. (2009), who have
satisfied the continuity of transverse Lagrangian displacement and
total pressure at each tube lateral surface to obtain $T^{j}$.
The expansion coefficient of order $m$ of the $j-$th loop,
$\alpha_m^j$, are coupled to all expansion coefficients of the
other loops, which reflects the collective nature of the normal
modes. For $N$ loops and $m_t$ expansion coefficients for each
field, there are $N\times(2m_t+1)$ equations from Eq.
(\ref{alpha}), whereas $m_t$ is the truncation number.
\subsection{Dispersion relations of longitudinal part, Eq. (\ref{Z})}
The boundary conditions, at footpoint, $z=0$, and apex, $z=L/2$,
for even and odd modes in the longitudinal direction are
\begin{eqnarray}\label{ex1}
\al\al \hspace{-1cm}Z(z=0)=Z(z=L/2)=0
~, \rm{even ~modes},\\
\al\al \hspace{-1cm}Z(z=0)=Z'(z=L/2)=0
~, \rm{odd ~modes}.\label{ex2}
\end{eqnarray}
Using Eq. (\ref{Z}) and imposing the boundary conditions, Eqs
(\ref{ex1}) and (\ref{ex2}), we can get the following dispersion
relations
\begin{eqnarray}\label{disp1}
\al\al \f{J_{\nu}(\alpha e^{-\epsilon/4 })}
{J_{\nu}(\alpha)}=
\f{Y_{\nu}(\alpha e^{-\epsilon/4 })}
{Y_{\nu}(\alpha)},~\rm{even ~modes},~~~~~\\\label{disp2}
\al\al \f{J'_{\nu}(\alpha e^{-\epsilon/4 })}
{J_{\nu}(\alpha)}=
\f{Y'_{\nu}(\alpha e^{-\epsilon/4 })}
{Y_{\nu}(\alpha)},~\rm{odd ~modes}.~~~~~~\\
\al\al \alpha=\f{2\tilde{\omega}}{\epsilon}.\nonumber
\end{eqnarray}

Solving the set of equations, Eq. (\ref{alpha}) and Eqs
(\ref{disp1}) and (\ref{disp2}), gives the frequency, $\omega$,
and the wave number, $k_z$, simultaneously. Equations
(\ref{disp1}) and (\ref{disp2}) are considered the effect of
longitudinal density stratification on collective transverse
oscillations, which is directly related to the main goal of the
present paper.
\section{Numerical results and conclusions}\label{numeric}
\subsection{Results for two similar loops}
The set of Eqs (\ref{alpha}), (\ref{disp1}), and (\ref{disp2})
are solved numerically for $\omega$ and $k_z$, based on
trust-region-dogleg algorithm. The dimensionless parameters are,
the frequencies, $\omega L/v_{A_{1}}(\epsilon=0)$, the
longitudinal wave number, $Lk_z$, the tube length scale, $a/L$,
the separation distance between center of loops, $d/a$, the
stratification parameter, $L/H$, the densities
$\rho_j(\epsilon)/\rho_0(\epsilon)$ and
$\rho_0(\epsilon)/\rho_0(0)$.

\begin{figure}
 \includegraphics{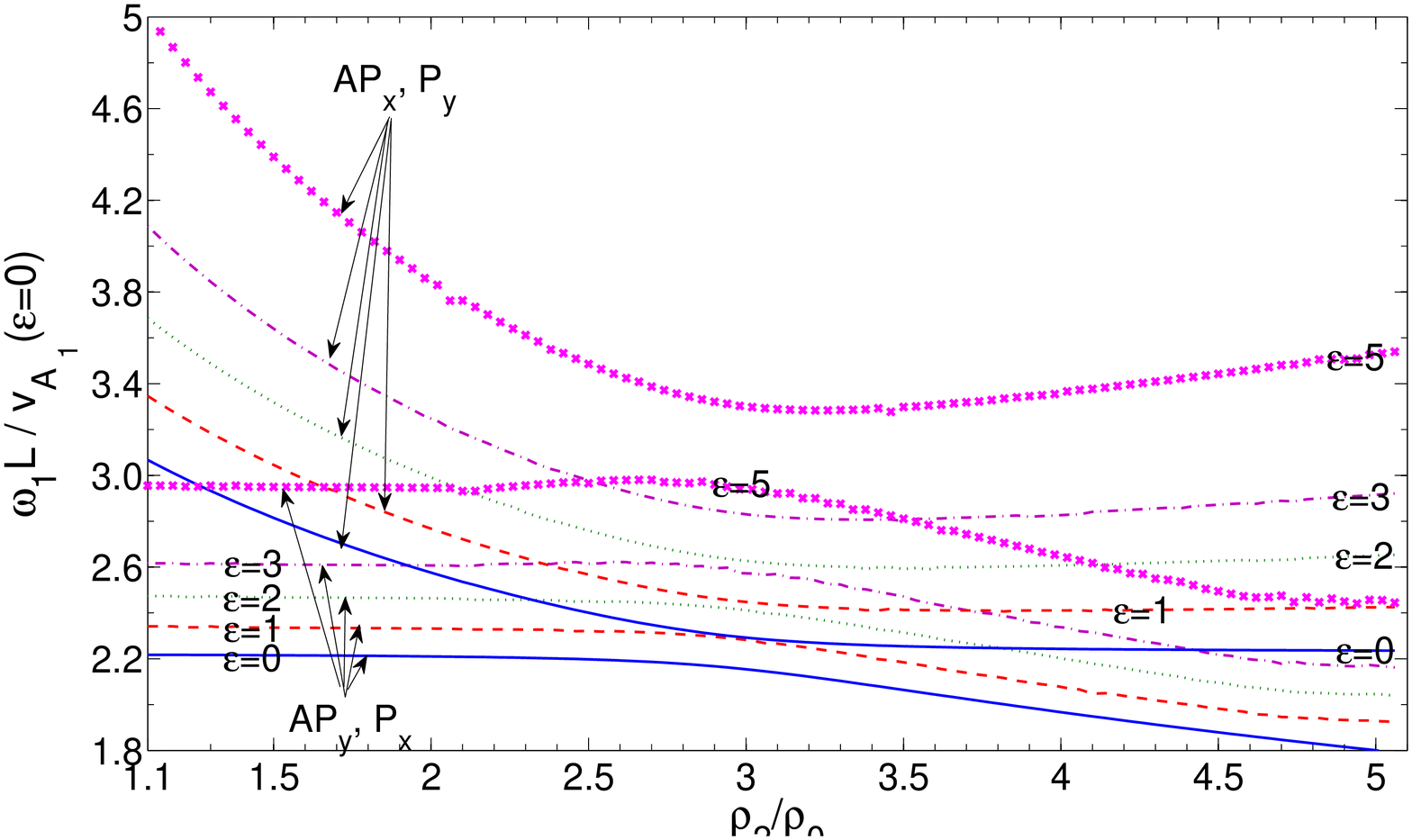}
         \vspace{4.5cm}
       \caption{The dimensionless fundamental frequencies, $\omega_1 L/v_{A_{1}}(\epsilon=0)$,
        plotted versus the density of the
second loop, $\rho_2/\rho_0$, and for different $\epsilon$
(solid lines $\epsilon=0$, dashed lines $\epsilon=1$,
dotted lines $\epsilon=2$, dash-dotted lines $\epsilon=3$, and cross lines $\epsilon=5$).
            }
            \label{fig1}
\end{figure}

Study the influence of the density on the normal mode properties,
we consider a system of two loops with radiuses $a_1 = a_2 = a =
0.03L$ with their centers separated a distance $d = 3a$, along
the $x$ axis. The first loop density is $\rho_{1} = 3\rho_0$
while $\rho_{2}$ is changed from $\rho_{2} = 1.1\rho_0$ to
$5\rho_0$. Just like Luna et al. (2009), we concentrate on the
kinklike modes which is the most important mode frequency of
transverse oscillations and
 find four kinklike normal
modes named $P_x$, $AP_y$, $P_y$, and $AP_x$, where $P$ and $AP$
refer to phase and antiphase motions of the loops, respectively.
In numerical processes, the truncation wave number, $m_t$, is cut
to $m=10$. Note that in the case of individual loop oscillations,
the kink modes are called with $m=1$. Our numerical results, Fig.
\ref{fig1}, show that, the discrepancy between $P_x$ and $AP_y$,
$P_y$ and $AP_x$, for stratified and unstratified system of loops
is less than $0.01$, which verified Luna et al. (2009) results.

In Fig. \ref{fig1}, dimensionless frequencies, $\omega
L/v_{A_{1}}(\epsilon=0)$, are plotted versus the second loop
density, $\rho_2/\rho_0$, and for different density stratification
parameter, $\epsilon$. As is shown in the figure: a) in the case
of unstratified system of coronal loops, $\epsilon=0$, our results
are fully in agreement with Luna et al. (2009), (see solid lines),
b) as $\epsilon$ increases the frequencies increase either, more
for $P_y$ and less for $P_x$ frequencies (dashed lines
$\epsilon=1$, dotted lines $\epsilon=2$, and dash-dotted lines
$\epsilon=3$), c) in density contrast, in the case of
$\rho_2=\rho_1$, frequencies discrepancy (i.e.,$ P_y-P_x$)
increases as $\epsilon$ increases.

\begin{figure}
 \includegraphics{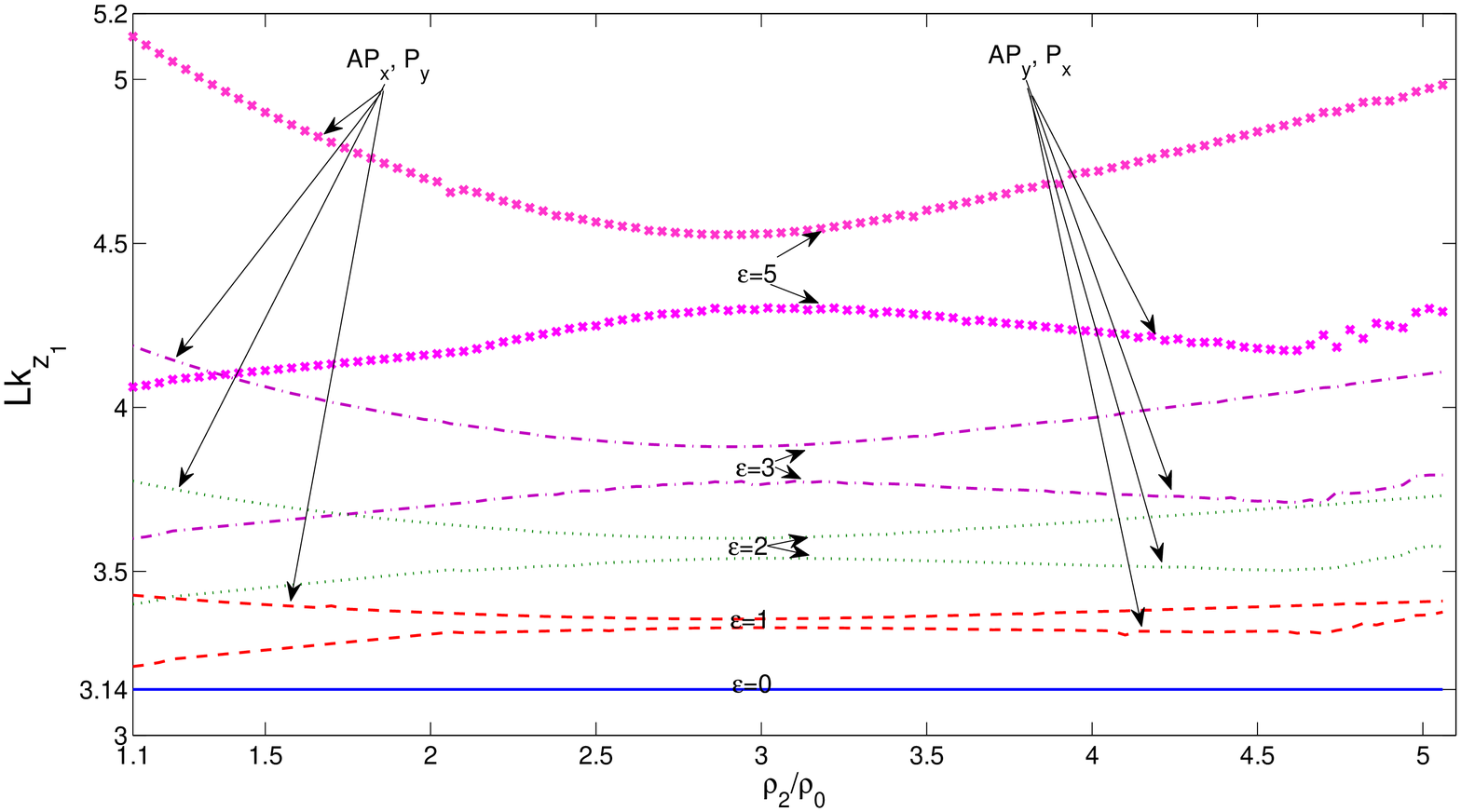}
         \vspace{5cm}
       \caption{Fundamental wave numbers, $Lk_{{z}_1}$, plotted versus $\rho_2/\rho_0$ and $\epsilon$
       (solid line $\epsilon=0$, dashed lines $\epsilon=1$, dotted lines $\epsilon=2$,
       dash-dotted lines $\epsilon=3$, and cross lines $\epsilon=5$).
            }
            \label{fig2}
\end{figure}

Figure \ref{fig2} shows dimensionless wave numbers, $Lk_z$, versus
the density of the second loop, $\rho_2/\rho_0$, and for different
density stratification parameter, $\epsilon$. The solid line shows
the wave numbers, for phase, $P_x$ and $P_y$, and antiphase,
$AP_x$ and $AP_y$ modes, for the case of $\epsilon=0$. We see
that, the wave numbers of these modes are degenerated, and as
$\epsilon$ increases the degeneracy is broken to two separated
branches (for $P_x$, $AP_y$ and $P_y$, $AP_x$ modes). Beside that,
$P_x$ and  $AP_y$ ($P_y$ and $AP_x$) themselves are closely
degenerated.  Luna et al. (2009) indicated that, the loop length
scale (the ratio of loop radius to loop length $a/L$) brokes this
degeneracy (see Fig. 3 therein).

\begin{figure}
 \includegraphics{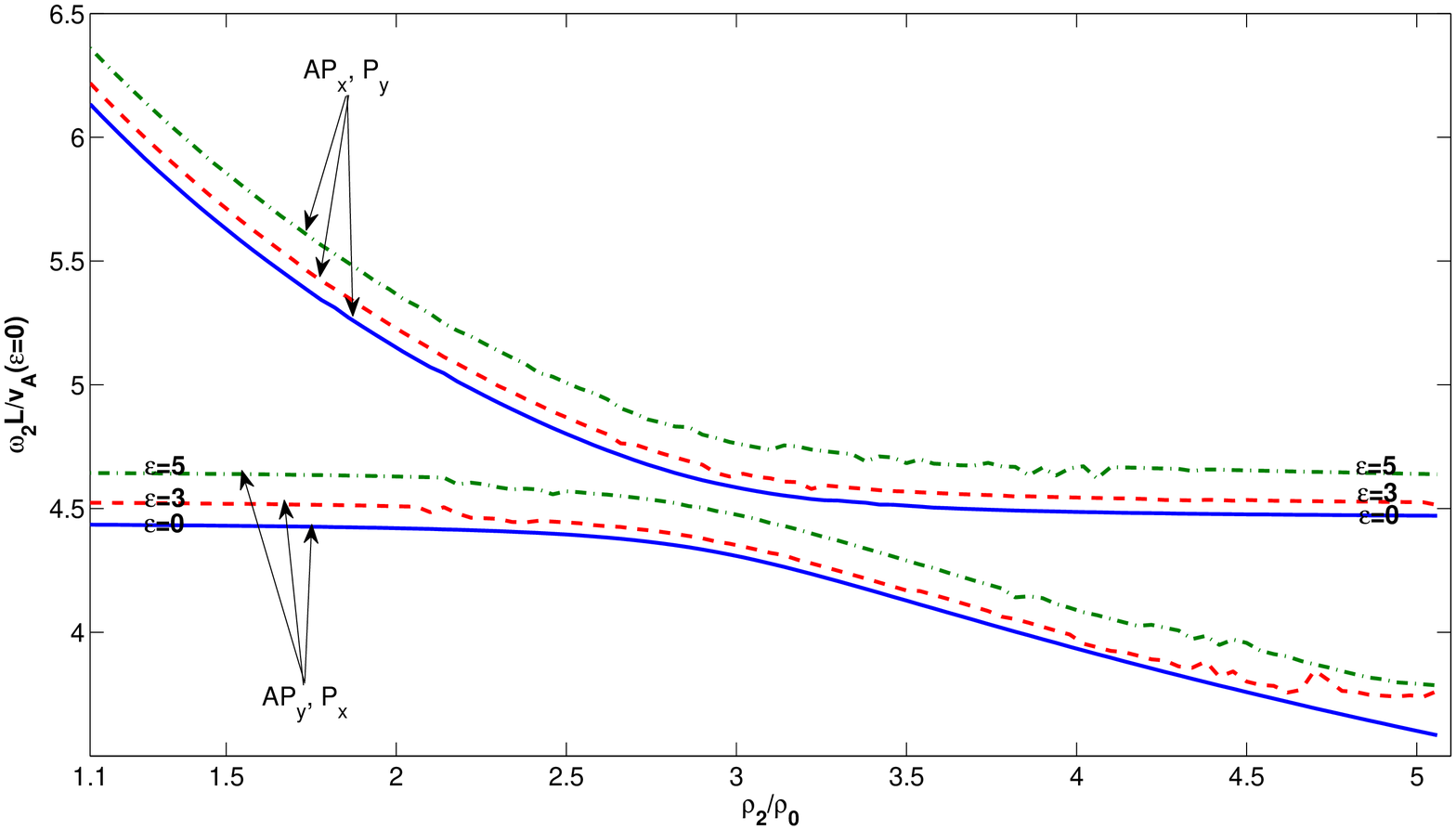}
         \vspace{5cm}
       \caption{The first dimensionless overtone frequencies, $\omega_2 L/v_{A_{1}}(\epsilon=0)$,
        plotted versus the density of the
second loop, $\rho_2/\rho_0$, and for different $\epsilon$
(solid lines $\epsilon=0$, dashed lines $\epsilon=3$,
dash-dotted lines $\epsilon=5$).
            }
            \label{fig3}
\end{figure}

In Fig. \ref{fig3}, the first dimensionless overtone frequencies
(the first even modes), $\omega_2 L/v_{A_{1}}(\epsilon=0)$, are
plotted versus the density of the second loop, $\rho_2/\rho_0$,
and for different density stratification parameter, $\epsilon$.
The first overtone frequencies, $\omega_2$, increase with
increasing of $\epsilon$ and decrease with increasing
$\rho_2/\rho_0$. Expectedly, as shown in Figs \ref{fig1} and
\ref{fig3}, for small stratification parameter ($\epsilon<3$), as
$\epsilon$ increases the first overtone frequencies, $\omega_2$,
increase more slightly than the fundamental frequencies,
$\omega_1$  (Safari et al. 2007 and Fathalian et al. 2010).

\begin{figure}
 \includegraphics{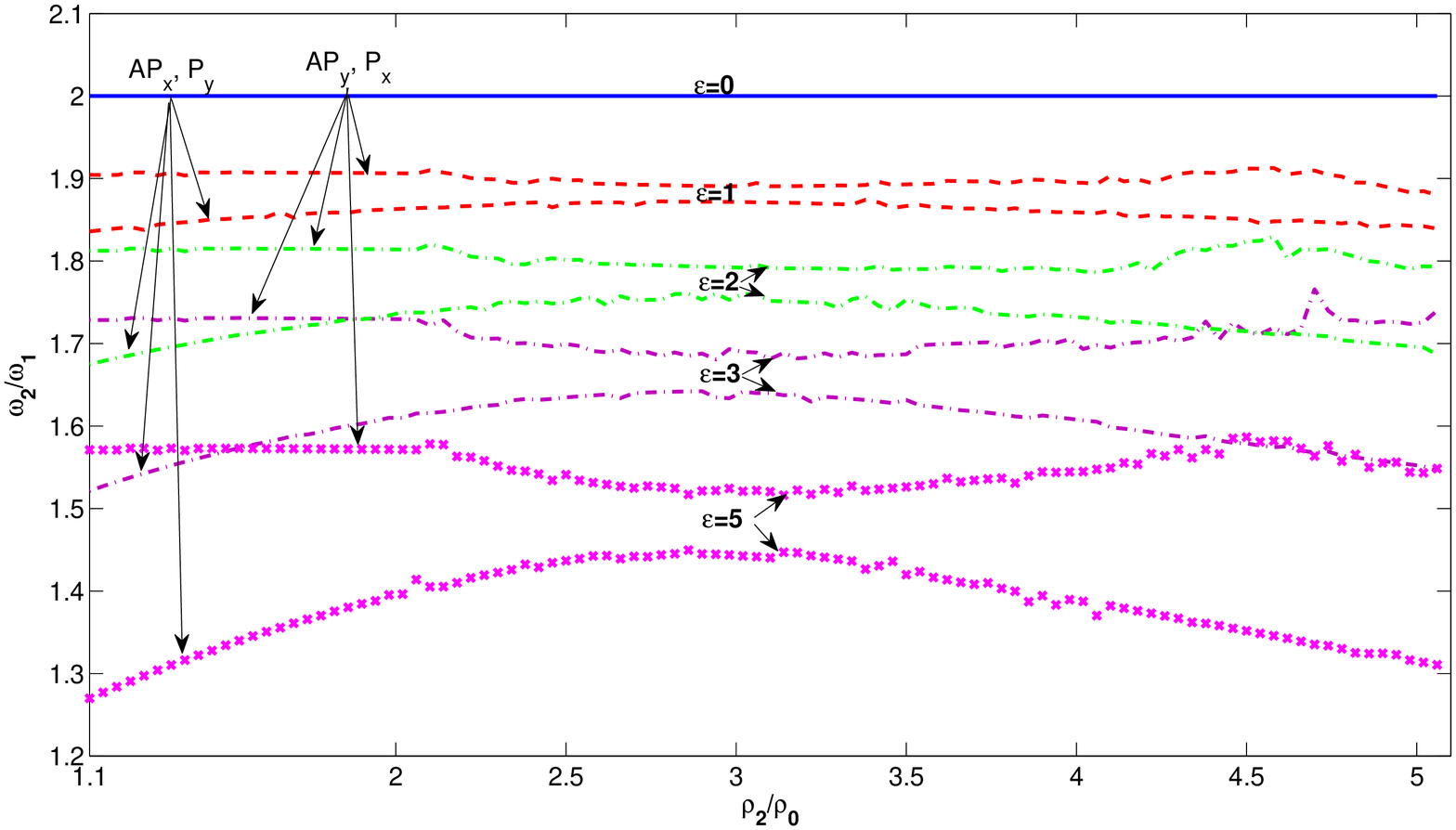}
         \vspace{5cm}
       \caption{The ratios of the first overtone frequencies to the fundamental frequencies,
$\omega_2/\omega_1$, are plotted versus the density of the
second loop, $\rho_2/\rho_0$, and for different $\epsilon$,
for $P_x$, $AP_y$ and $P_y$, $AP_x$ modes (solid line $\epsilon=0$, dashed lines $\epsilon=1$,
dotted lines $\epsilon=2$, dash-dotted lines $\epsilon=3$, and cross lines $\epsilon=5$).
}
            \label{fig5}
\end{figure}

In Fig.\ref{fig5},
the ratios of the first overtone frequencies to the fundamental frequencies,
$\omega_2/\omega_1$ (for $P_x$, $AP_y$ and $P_y$, $AP_x$ modes) are plotted versus $\rho_2/\rho_0$
 and for different $\epsilon$.
As was expected, all of the ratios are less than 2 for stratified
cases and
  decreased by increasing $\epsilon$. In the case of unstratified
   system of coronal loops, $\epsilon=0$,
  the ratios of phase and antiphase frequencies are degenerated.
As we see, in the case of $\epsilon=0$, the frequencies ratios
don't change with changing of density, $\rho_2/\rho_0$. This means
that, the frequencies ratios of unstratified system of coronal
loops are the same as monoloop ones.
 As $\epsilon$ increases this degeneracy is broken to
two separated branches (for $P_x$, $AP_y$ and $P_y$, $AP_x$).

\begin{figure}
 \includegraphics{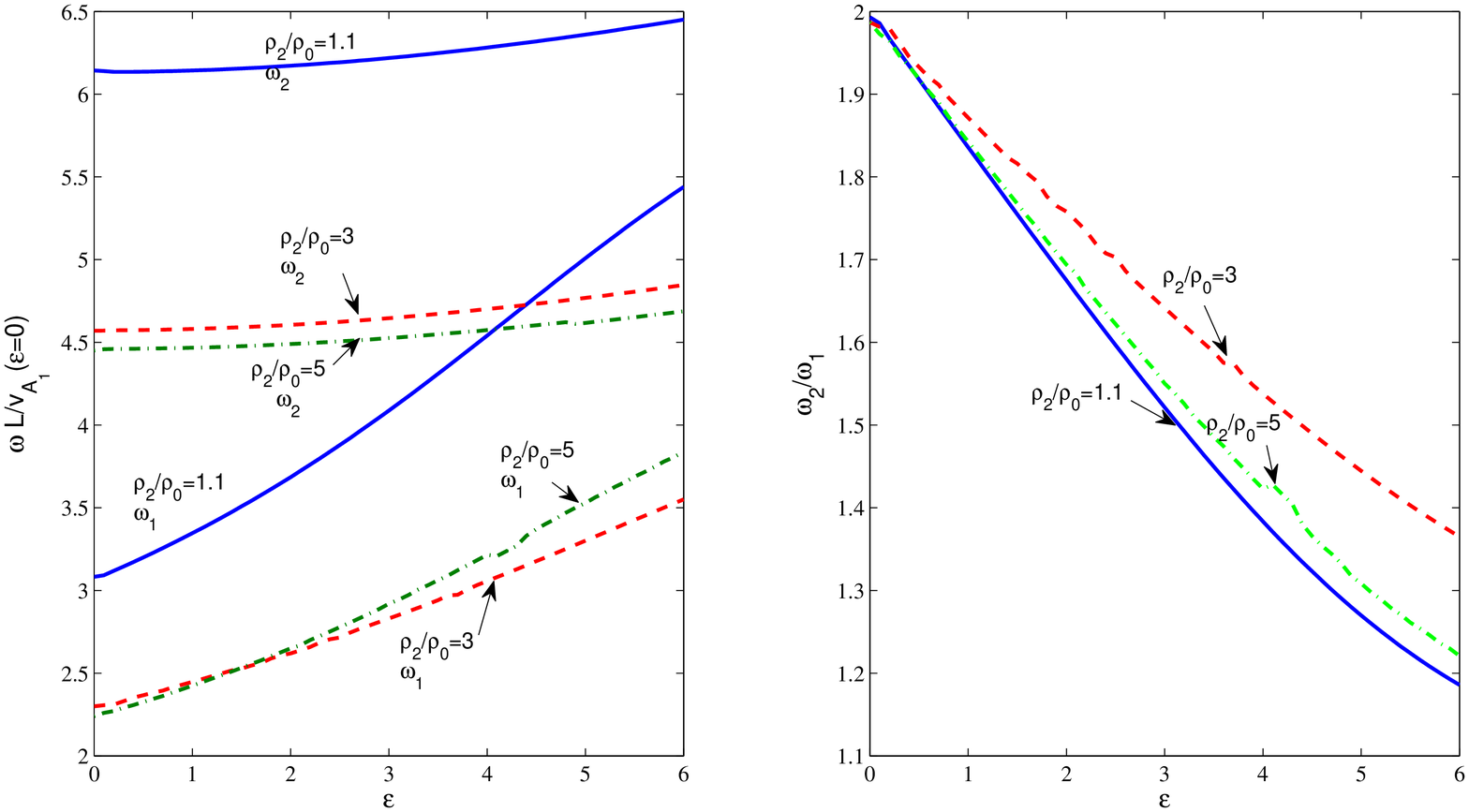}
         \vspace{6cm}
       \caption{The dimensionless frequencies, $\omega L/v_{A_{1}}(\epsilon=0)$,
       and the frequencies ratios, $\omega_2/\omega_1$,
 are plotted as a
 function of $\epsilon$ and  different $\rho_2$.
       The first loop density is $\rho_{1} = 3\rho_0$.
The solid, dashed, and dash-dotted lines are for
$\rho_{2}=1.1\rho_0$, $\rho_{2}=3\rho_0$, and $\rho_{2}=5\rho_0$,
for ($AP_{x}$, $P_y$). We see that with increasing of $\epsilon$
the frequencies increase expectedly.
            }
            \label{fig6}
\end{figure}

In Fig.\ref{fig6}, dimensionless frequencies, $\omega
L/v_{A_{1}}(\epsilon=0)$, and frequencies ratios,
$\omega_2/\omega_1$,
 are plotted as a
 function of $\epsilon$ and  different $\rho_2$.
 The first loop density is $\rho_{1} = 3\rho_0$.
The solid, dashed, and dash-dotted lines are for
$\rho_{2}=1.1\rho_0$, $\rho_{2}=3\rho_0$, and $\rho_{2}=5\rho_0$,
for ($AP_{x}$, $P_y$). Expectedly, in the case of
$\rho_{2}=1.1\rho_0$, the frequencies ratios, $\omega_2/\omega_1$,
is close to the monoloop frequencies ratio (solid line, Fig.
\ref{fig6}).
 We see that
with increasing of $\epsilon$ the frequencies increase.
Frequencies ratio, in the case of $\rho_{2}=3\rho_0$, which is
equal to the first loop density, differs from the other ones.

Van Doorsselaere et al. (2007) revisited observational
frequencies ratios of Verwichte et al. (2004), to be in the range
of $1.58\leq \omega_2/\omega_1 \leq 1.82$. Using Fig. \ref{fig6},
and for typical loop length, $100$Mm, corresponding values of the
density scale height, $H=\epsilon^{-1}(L)$, fall in the range of
$[38, 91]$Mm for $\rho_2=1.1\rho_0$, $[36, 87]$Mm for
$\rho_2=5\rho_0$, and $[28, 69]$Mm for the case of equal density
of two loops $\rho_1=\rho_2=3\rho_0$. We see that, for a system
of two loops with neighboring similar densities, the estimated
density scale height changed significantly.
 Observations with
higher resolutions such as Solar Dynamics Observatory could
improve the practical values of frequencies ratio. At that stage,
we can compare the effects of different factors (e.g., non-uniform
magnetic flux tubes with variable cross sections, system of many
loops or multi-strands loops, etc) on the loops dynamics. In the
present stage of the three observations, Van Doorsselaere et al.
(2007), we can only be encouraged as our expectation of a density
scale height is around 50-100 Mm.

\subsection{Conclusions}\label{con}
In this paper, we extended the study of collective transverse
oscillations of system of coronal loops with both radial and
longitudinal density stratification. To do this, we composed two
different approaches, Luna et al. (2009) approach (used for
studying the kinklike oscillations of system of coronal loops) and
Safari et al. (2010) approach (applied for studying  single
isolated stratified coronal loop). Our main results are listed
briefly as:
\begin{description}
\item[-] In the presence of stratification parameter, $\epsilon$,
we can get dispersion relations (Eqs. \ref{disp1} and
\ref{disp2}), and solve them
 numerically for the fundamental and first overtone phase and antiphase
 modes.
 Similar as Luna et al. (2009), we focused on the collective
 kinklike oscillations. In stratified system of coronal loops, we
 see that the frequencies of kinklike normal modes are more or less as
 kink.
\item[-] Density stratification changes the fundamental and first
overtone kinklike frequencies and their ratios for systems of two
coronal loops. The order of changes for the frequency ratios of
two coupled loops in respect of one loop, is about ${\cal O}$$
(10^{-4})$, and longitudinally stratification has more effect on
frequencies ratios with decreasing and degenerating the ratios
for phase and antiphase modes.
\item[-]
 The frequencies ratios of unstratified system of coronal
loops ($\epsilon=0$), are the same as for monoloop.
Longitudinally density stratification is completely important in
dynamics of system of coronal loops, which breaks the existent
degeneracy to two separated branches (pairs $P_x$, $AP_y$ and
$P_y$, $AP_x$).
\end{description}
The difficulties in the numerical solutions come from
discrimination of different mode numbers with close frequency
values. To avoid this, double precision was used in our numerical
processes.
\section*{Acknowledgments}
 The authors would like to thank the unknown referee for his/her very helpful comments and
   suggestions.

\clearpage

\begin{thebibliography}{}
\bibitem{Andries2005a}Andries, J., Arregui, I., \& Goossens, M. 2005a, ApJ 624, L57.
\bibitem{Andries2005b}Andries, J., Goossens, M., Hollweg, J. V., Arregui, I., \& Van
Doorsselaere, T. 2005b, A\&A 430, 1109.
\bibitem{Andries2009}Andries, J., Van Doorsselaere, T., Roberts, B., Verth, G., Verwichte, E., \& Erd\'{e}lyi,
R. 2009, Space Sci. Rev., 149, 3.
\bibitem{Aschwanden2009}Aschwanden, M. J. 2009, Space Sci. Rev., 149,
31.
\bibitem{Aschwanden2002}Aschwanden, M. J., De Pontieu, B., Schrijver, C. J., \& Title, A. M. 2002, Sol. Phys., 206, 99.
\bibitem{Aschwanden1999a}Aschwanden, M. J., Fletcher, L., Schrijver, C. J., \& Alexander, D. 1999a, ApJ, 520, 880.
\bibitem{Aschwanden1999b}Aschwanden, M. J., Newmark, J. S., Delaboudini\'{e}re, J., Neupert, W.
M., \&  Klimchuk, J. A., et al. 1999b, ApJ, 515, 842.
\bibitem{De Moortel2007}De Moortel, I. \& Brady, C. S. 2007, ApJ, 664, 1210.
\bibitem{Diaz2005}D\'{i}az, A. J., Oliver, R., \& Ballester, J. L. 2005, A\&A, 440, 1167.
\bibitem{Donnelly2006}Donnelly, G. R., D\'{i}az, A. J., \& Roberts, B., 2006. A\&A, 457,
707.
\bibitem{Dymova2006}Dymova, M. V. \& Ruderman, M. S. 2006, A\&A,
459, 241.
\bibitem{Edwin} Edwin P. M. \& Roberts, B. 1983, Sol. Phys., 88, 179.

\bibitem{Erdelyi2007}Erd\'{e}lyi, R. \& Verth, G. 2007, A\&A, 462,
743.
\bibitem{Fathalian2010}Fathalian, N., Safari, H., \& Nasiri, S. 2010, New Astronomy, 15, 403.
\bibitem{Goossens92}Goossens, M., Hollweg, J. V., \& Sakurai, T. 1992, Sol. Phys., 138, 233.
\bibitem{Gruszecki2006}Gruszecki, M., Murawski, K., Selwa, M., \& Ofman,
L. 2006, A\&A, 460, 887.
\bibitem{Klimchuk2006}Klimchuk, J. A. 2006, Sol. Phys., 234, 41.
\bibitem{Luna2009}Luna, M., Terradas J., Oliver R., \& Ballester J. L. 2009, ApJ, 692, 1582.
\bibitem{Luna2010}Luna, M., Terradas, J., Oliver, R., \& Ballester, J.
L. 2010, ApJ, 716, 1371.
\bibitem{McEwan2006}McEwan, M., Donnelly, G.R., D\'{i}az, A.J., \&  Roberts, B. 2006,
A\&A, 460, 893.
\bibitem{Murawski93}Murawski, K. 1993, Acta Astronomica, 43, 2, 161.
\bibitem{Murawski94}Murawski, K. \& Roberts, B. 1994, Sol. Phys., 151, 305.
\bibitem{Nakariakov1999}Nakariakov, V. M., Ofman, L., DeLuca, E. E., Roberts, B., \& Davila, J. M. 1999, Science,
285, 862.
\bibitem{Pascoe2007}Pascoe, D. J., Nakariakov, V. M., \&  Arber, T.
D. 2007, Sol. Phys., 246, 165.
\bibitem{Ramm86}Ramm, A. G. 1986, Scattering by obstacles (Dordrecht: Reidel).
\bibitem{Roberts84}Roberts, B., Edwin, P. M., \& Benz, A. O. 1984, ApJ, 279, 857.
\bibitem{Ruderman2008}Ruderman, M.S., Verth, G., \& Erd\'{e}lyi, R. 2008, ApJ, 686, 694.
\bibitem{Safari2007}Safari, H., Nasiri, S., \& Sobouti Y. 2007, A\&A, 470, 1111.
\bibitem{Schrijver.Brown2000}
Schrijver, C. J. \& Brown, D. S. 2000, ApJ, 537, L69.
\bibitem{Doorsselaere}Van Doorsselaere, T., Nakariakov, V. M., \& Verwichte, E. 2007, A\&A, 473,
959.
\bibitem{Verwichte2004}
Verwichte, E., Nakariakov, V. M., Ofman, L., \& Deluca, E. E. 2004, Sol. Phys., 223, 77.
\bibitem{Waterman1961}Waterman, P. C. \& Truell, R. 1961, JMP, 2, 512.
\end{thebibliography}
\end{document}